\documentclass[conference,10pt]{IEEEtran}
\normalsize
\usepackage{bbm}
\usepackage{adjustbox}
\usepackage{enumitem}
\usepackage{cite,algorithm,algorithmic,amsmath,amssymb,amsthm,empheq,mhsetup}
\usepackage{subfigure,amsfonts,balance}
\usepackage{epstopdf}
\usepackage{enumitem}
\usepackage{setspace}
\usepackage[dvipsnames]{xcolor}
\usepackage[left=0.649in, right=0.68in, top=0.71in, bottom=1.049in,centering]{geometry}
\usepackage{lipsum} % For some dummy text
\usepackage{filecontents} % To make the bib-file

\DeclareMathOperator{\EEE}{\mathbb{E}}

\DeclareMathOperator{\f}{\pmb{f}}

\DeclareMathOperator{\FF}{\mathcal{F}}

\DeclareMathOperator{\K}{\mathcal{K}}

\DeclareMathOperator{\OO}{\mathcal{O}}

\DeclareMathOperator{\vv}{\pmb{v}}

\DeclareMathOperator{\G}{\pmb{G}}

\DeclareMathOperator{\CN}{\mathcal{CN}}

\DeclareMathOperator{\NN}{\mathcal{N}}

\DeclareMathOperator{\e}{\pmb{e}}

\DeclareMathOperator{\rr}{\pmb{r}}
\DeclareMathOperator{\x}{\pmb{x}}

\DeclareMathOperator{\uu}{\pmb{u}}

\DeclareMathOperator{\q}{\pmb{q}}
\DeclareMathOperator{\g}{\pmb{g}}

\DeclareMathOperator{\OOmega}{\pmb{\omega}}
\DeclareMathOperator{\ETA}{\pmb{\eta}}

\DeclareMathOperator{\THeta}{\pmb{\theta}}

\DeclareMathOperator{\ZETA}{\pmb{\zeta}}

\newtheorem{remark}{Remark}

\ifCLASSINFOpdf
\else
\fi

\makeatletter
\def\bstctlcite{\@ifnextchar[{\@bstctlcite}{\@bstctlcite[@auxout]}}
\def\@bstctlcite[#1]#2{\@bsphack
	\@for\@citeb:=#2\do{%
		\edef\@citeb{\expandafter\@firstofone\@citeb}%
		\if@filesw\immediate\write\csname #1\endcsname{\string\citation{\@citeb}}\fi}%
	\@esphack}
\makeatother

\begin{document}
	\bstctlcite{IEEEexample:BSTcontrol}
	\fontsize{10}{12}\rm
	\title{\Huge  Energy-Efficient Massive MIMO for Serving Multiple Federated Learning Groups\vspace{-4mm}}
	\author{
		\IEEEauthorblockN{
			Tung T. Vu\IEEEauthorrefmark{1},
			Hien Quoc Ngo\IEEEauthorrefmark{1},
			Duy T. Ngo\IEEEauthorrefmark{2},
			Minh N Dao\IEEEauthorrefmark{3},
			Erik G. Larsson\IEEEauthorrefmark{4}
			%Duy T. Ngo\IEEEauthorrefmark{1},
			%Minh N. Dao\IEEEauthorrefmark{3}
			%Nguyen H. Tran\IEEEauthorrefmark{4},
			%and Richard H. Middleton\IEEEauthorrefmark{1}
			%Minh N. Dao\IEEEauthorrefmark{1},
			%Salman Durrani\IEEEauthorrefmark{4},
			%and Richard H. Middleton\IEEEauthorrefmark{1}
		}
		\IEEEauthorblockA{\small\IEEEauthorrefmark{1}Institute of Electronics, Communications, and Information Technology (ECIT), Queen's University Belfast, Belfast BT3 9DT, UK}
		\IEEEauthorblockA{\small\IEEEauthorrefmark{2}School of Electrical Engineering and Computing, The University of Newcastle, Callaghan, NSW 2308, Australia}
		\IEEEauthorblockA{\small\IEEEauthorrefmark{3}School of Engineering, Information Technology and Physical Sciences, Federation University, Ballarat, VIC 3353, Australia}
		\IEEEauthorblockA{\small\IEEEauthorrefmark{4}Department of Electrical Engineering (ISY), Link\"{o}ping University, SE-581 83 Link\"{o}ping, Sweden}
		\IEEEauthorblockA{
			%\small\IEEEauthorrefmark{3}School of Engineering, Information Technology and Physical Sciences, Federation University Australia, Ballarat, VIC 3353, Australia
			%\\
			Email: t.vu@qub.ac.uk, hien.ngo@qub.ac.uk, %m.matthaiou@qub.ac.uk, tom.marzetta@nyu.edu
			duy.ngo@newcastle.edu.au, m.dao@federation.edu.au,
			%, nguyen.tran@sydney.edu.au, 
			erik.g.larsson@liu.se
		}
		%\author{
			%\IEEEauthorblockN{Author 1 and Author 2}
			%\IEEEauthorblockA{\small\\
				% \\
				%    }
			%\thanks{This work is supported in part by an ECR-HDR scholarship from The University of Newcastle and in part by an Australian Research Council Discovery Project grant DP170100939.}
			\vspace{-10mm}
		}
		
		\maketitle
		\allowdisplaybreaks
		\begin{spacing}{1}
			\begin{abstract}
				With its privacy preservation and communication efficiency, federated learning (FL) has emerged as a learning framework that suits beyond 5G and towards 6G systems. This work looks into a future scenario in which there are multiple groups with different learning purposes and participating in different FL processes. We give energy-efficient solutions to demonstrate that this scenario can be realistic. First, to ensure a stable operation of multiple FL processes over wireless channels, we propose to use a massive multiple-input multiple-output network to support the local and global FL training updates, and let the iterations of these FL processes be executed within the same large-scale coherence time. Then, we develop asynchronous and synchronous transmission protocols where these iterations are asynchronously and synchronously executed, respectively, using the downlink unicasting and conventional uplink transmission schemes. Zero-forcing processing is utilized for both uplink and downlink transmissions. Finally, we propose an algorithm that optimally allocates power and computation resources to save energy at both base station and user sides, while guaranteeing a given maximum execution time threshold of each FL iteration. Compared to the baseline schemes, the proposed algorithm significantly reduces the energy consumption, especially when the number of base station antennas is large.
			\end{abstract}
		\end{spacing}
		
		\vspace{-0mm}
		% \begin{IEEEkeywords}
			% Energy consumption, massive MIMO, multiple federated learning groups, zero-forcing.
			% \end{IEEEkeywords}

		% \fi
		\IEEEpeerreviewmaketitle
		
		%\balance
		\vspace{-3mm}
		\section{Introduction}
		\vspace{-1mm}
		\label{sec:Introd}
		Recently, federated learning (FL) was introduced in \cite{mcmahan17AISTATS} as an important step to bring machine learning closer to everyone. The breakthrough idea of FL is ``no raw data sent to third party companies during learning processes'', which means people can safely participate in FL processes without being worried that their personal data is exploited. 
		A wide range of applications, such as healthcare and self-driving cars to name  a few \cite{chen20IS,niknam20CM}, can benefit from  FL. 
		In FL, the learning is implemented jointly by many users (UEs). First, a local learning model is trained at each UE using local (private) training data, and sent to the central server. A global update is then computed at the central server using the local learning models transmitted from all UEs, and finally sent back to the UEs for local training updates. This learning process is iterated  until reaching a certain learning accuracy level. 
		To deploy the above iterative FL process over wireless networks, a key challenge is keeping the network energy consumption as low as possible. This is important both due to  battery limitations  of the UEs and to   concerns about the ICT carbon footprint. It is thus critical to design an energy-efficient wireless network to support FL.  
		
		There are several studies of  energy-efficient deployments of FL over wireless networks; see, e.g.,   \cite{yang21TWC,zeng20ICC,hu20WCSP} and references therein. In these works, the authors proposed designs which  minimize the energy consumption at the UEs  while guaranteeing the learning performance (i.e., the test accuracy) by jointly optimizing learning and communication parameters. However, the energy consumption of the transmission from the central server to the UEs was not taken into account. Also, these works proposed to use frequency-division multiple access (FDMA) and time-division multiple access (TDMA) systems  to support FL. This might not be a good choice because FDMA and TDMA systems offer low UE data rates, and hence, yield a very high energy consumption, especially when the number of UEs is large. In addition, all these works only considered the case of a single FL group. 
		
		On the other hand, it is anticipated that the future wireless systems will need to serve multiple groups of UEs that  participate in  different FL processes.  These networks need to simultaneously provide high data rates and high communication reliability to all UEs in all FL groups. Designing such networks is challenging and calls for a suitable, new wireless communication framework. To the best of our knowledge, there has not been any work studying energy-efficient wireless networks supporting multiple FL groups in the existing literature.
		
		% \textit{Paper Contribution:}
		The contributions of this paper are summarized as follows:
		\vspace{-2mm}
		\begin{itemize}[noitemsep,nolistsep]
			\item To support multiple FL groups over wireless networks, we  propose using massive MIMO (mMIMO) and letting multiple iterations (each for one FL process of a group) be executed in one large-scale coherence time\footnote{The large-scale coherence time is defined as the time interval during which  the large-scale fading coefficients remain approximately constant.}. Thanks to the high array gain and multiplexing gain, mMIMO can simultaneously offer very high quality of service to all UEs in an area of interest \cite{ngo16}, and hence, it is expected to guarantee a stable operation of each iteration (and hence the whole FL process).
			\item We introduce two specific transmission protocols where the steps within one FL iteration, i.e., the downlink transmission, the computation at the UEs, and the uplink transmission, are either asynchronous or synchronous. These schemes   differ from the scheme in \cite{vu20TWC} which focuses on minimizing the training time of FL. Here, we use the unicast protocol on  downlink and   conventional multiuser transmission on uplink.
			Both downlink and uplink use zero-forcing (ZF) processing. 
			\item We develop an algorithm to allocate the transmit powers and processing frequencies to minimize the total energy consumption in each FL iteration, under a constraint on the total  time taken for one FL iteration.
			\item Numerical results show that our proposed schemes significantly reduce the energy consumption compared to baseline schemes. They also confirm that the asynchronous scheme outperforms the synchronous scheme for supporting multiple FL groups, at the cost of a higher complexity.
		\end{itemize}

		\vspace{-2mm}
		\section{Proposed Schemes and System Model}
		\label{sec:SystModel}
		\vspace{-1mm}
		\subsection{Multiple Federated Learning Framework} \label{sec:FLframework}
		\vspace{-1mm}
		We consider a multiple FL network which includes multiple FL groups with different learning purposes. 
		Each UE is assumed to only participate in one FL group. The FL frameworks of all groups can be different in terms of loss functions but have the same following four steps in each iteration \cite{tran19INFOCOM,mcmahan17AISTATS}. 
		\begin{enumerate}[label={(S\arabic*)}]
			\item A central server sends a global update to the UEs.
			\item Each UE updates and solves its local learning problem using its local data and then computes its local update.
			\item Each UE sends its computed local update to the central server.
			\item The central server computes the global update by aggregating the received local updates from all UEs.
		\end{enumerate}
		The above process will be done iteratively until a certain learning accuracy level is achieved.
		
		\vspace{-2mm}
		\subsection{Proposed Schemes to Serve Multiple FL Groups}
\begin{figure}[t!]
			\centering
			\includegraphics[width=0.42\textwidth]{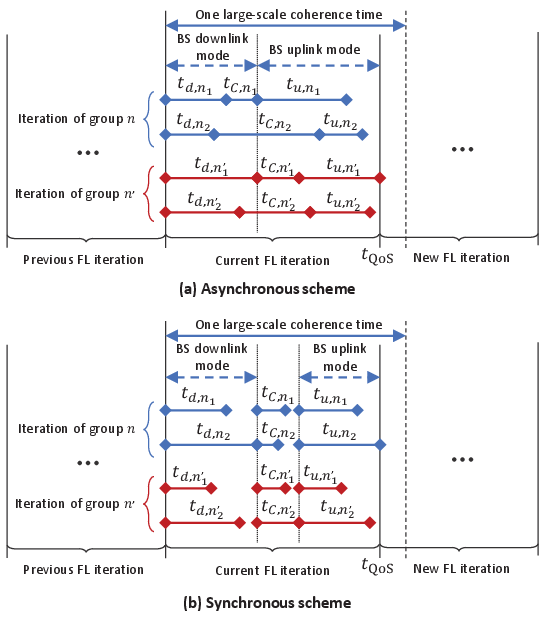}
			\vspace{-3mm}
			\caption{Illustration of FL iterations over the considered mMIMO network with two groups $n,n'$ and two UEs for each group}
			\label{fig:time1}
		\end{figure}

		\vspace{-1mm}
		To support multiple FLs  discussed in Section~\ref{sec:FLframework}, we propose to use mMIMO technology, i.e. Steps (S1) and (S3) of each FL iteration can be executed via the downlink and the uplink of a mMIMO system, respectively. Our proposed mMIMO-based multiple-FL system includes one $M$-antenna base station (BS) simultaneously serving $N$ FL groups in the same frequency bands under the time-division-duplexing operation. We assume that the BS acts as the central server. Each FL iteration of each FL group is assumed to be executed within a large-scale coherence time, which is reasonable because the execution time of one FL iteration is smaller than the large-scale coherence time in many scenarios \cite{vu20TWC}.
		With this assumption, we then propose two specific transmission schemes to support the learning of $N$ FL groups for each FL iteration as shown in Figs.~\ref{fig:time1}(a) and~(b) respectively.
		\vspace{-1mm}
		\begin{itemize}
			\item[(a)]\textbf{Asynchronous scheme}: All groups start their FL iterations at the same time when the BS switches to a downlink mode. During this mode, BS simultaneously sends the global updates to all UEs in all groups (corresponding to Step~(S1)). Each UE will start its local computation if it successfully receives the global training update (corresponding to Step~(S2)). Then, the BS switches to an uplink mode immediately after the receptions of the global training update are completed at all the UEs. During this mode, the UEs will send their computed local updates to the BS (corresponding to Step~(S3)) if they finish the local computation. 
			\item[(b)]\textbf{Synchronous scheme}: This scheme is similar to the asynchronous scheme except for the synchronization of Steps (S1)-(S3) among all the UEs.
			Each UE starts and waits for others to end each of those steps together.
		\end{itemize}
		The time of one FL iteration under both schemes are constrained by a given period of time.
		Note that in the asynchronous scheme, the time of Steps (S1)--(S3) are optimally allocated (using the the proposed algorithm in the next section) to ensure that all the UEs finish one FL iteration and start a new FL iteration at the same time.

		\vspace{-2mm}
		\subsection{Massive-MIMO-based Multiple-FL System Model}
		\vspace{-1mm}
		
		The above two schemes share the common system model as follows. 
		% As shown in Fig.~\ref{fig:time2}, 
		In each large-scale coherence time, the global and local updates in Steps (S1) and (S3) are transmitted in one or multiple small-scale coherence times depending on their sizes.
		Each coherence block in Step (S1) (or (S3)) involves the channel estimation phase and the downlink (or uplink) payload data phase.
		Suppose that at the considered time, there are $N$ iterations of $N$ FL groups being served. 
		Let 
		% $\MM\triangleq\{1,\dots,M\}$,
		$\NN\triangleq\{1,\dots,N\}$, and $\K_n$ be the sets of 
		% antennas, 
		groups and the indices of the UEs in group $n$, respectively. There are $K_n$ single-antenna users (UEs) of each group $n$. 
		The details of each step are presented in the following.

		% \begin{remark}
			% The channel estimation of the multicast scheme is worse than that of the unicast counterpart. This is because the multicast scheme introduces the pilot contamination within the UEs within one group, while the unicast scheme does not.
			% Inspite of that, the multicast scheme requires a shorter pilot length than the the unicast scheme. 
			% \end{remark}

		\subsubsection{Step (S1)}
		The BS sends the global updates to all UEs of all groups. Since the global updates intended for all UEs in a given group are the same, the transmission in this step corresponds to multi-group multicasting. Thus, we follow the scheme in \cite{sadeghi18TWC} assuming orthogonal pilots and ZF processing.  
		% Each coherence block of this step involves 2 main phases: uplink channel estimation and downlink unicasting.
		% Since the multicast estimation is performed, we choose a multicast transmission for this step because multicasting results in a higher performance due to a weaker pilot contamination in comparison with unicasting. 
		% On the other hand, if the unicast estimation is performed, we use the unicast transmission for this step because unicasting causes weaker inter-group interferences than unicating. Since we only consider these two cases, for ease of prepsentation, we simply name them as downlink multicasting and unicasting in this step.
		
		% \begin{itemize}
			% \item \textbf{Uplink channel estimation}:
			% \end{itemize}
		\textbf{Uplink channel estimation}:
		For each coherence block of length $\tau_c$, each UE sends its pilot of length $\tau_{d,p}$ to the BS \cite{sadeghi18TWC}. 
		We assume that the pilots of all the UEs are pairwisely orthogonal, which requires $\tau_{d,p}\geq K_{total}\triangleq \sum_{n\in\NN}K_n$. 
		% In practice, $N$ is relatively small, and hence, this requirement is normally satisfied. 
		% Let $\sqrt{\tau_p}\VARPHI_{n}\in\C^{\tau_p\times 1}$ be the same pilot sequence transmitted from $K_n$ UEs of group $n$, where $\|\VARPHI_{n}\|^2\!\!=\!\!1,\forall n\in\NN$.
		Denote by $\g_{n_k} \!=\! (\beta_{n_k})^{1/2}\tilde{\g}_{n_k}$ the channel vector from UE $k$ of group $n$ to the BS, where $\beta_{n_k}$ and $\tilde{\g}_{n_k}$ 
		% $\sim \CN(\pmb{0},\pmb{I}_M)$ 
		are the large-scale fading coefficient and small-scale fading coefficient vector, respectively.
		%Denote by $\g_n \triangleq \sum_{n_k\in\K_n}\sqrt{\tau_{cp} \rho_p} \g_{n_k}$ a linear combination of the channels of all the UEs within group $n$. 
		At the BS, $\g_{n_k}$ is estimated by using the received pilots and the minimum mean-square error (MMSE) estimation technique. The MMSE estimate $\hat{\g}_{n_k}$ of $\g_{n_k}$ is
		% $\hat{\g}_{n_k}=\hat{\sigma}_{n_k}\z_{mn}$ 
		distributed according to $\CN(\pmb{0},\hat{\sigma}_{n_k}^2\pmb{I}_M)$, where $\hat{\sigma}_{n_k}^2 = \frac{\tau_{d,p} \rho_{p} \beta_{n_k}^2 }{ \tau_{d,p} \rho_{p} \beta_{n_k} +1 }$ \cite{sadeghi18TWC}.
		% while the estimate of $\g_n$ is $\hat{\g}_n \sim \CN(\pmb{0}, \hat{\sigma}_{n}^2\pmb{I}_M)$, where $\hat{\sigma}_{n}^2 = \frac{(\tau_{cp} \rho_{p}\sum_{n_\ell \in \K_n} \beta_{n_\ell})^2 }{ \tau_{cp} \rho_{p} \sum_{\ell\in\K_n}  \beta_{n_\ell} +1 }$ \cite{sadeghi18TWC}. 
		We also denote by $\hat{\G}\triangleq [\hat{\G}_1,\dots,\hat{\G}_N]$ the matrix stacking the channels of all the UEs, where $\hat{\G}_n\triangleq [\hat{\g}_{n_1},\dots,\hat{\g}_{n_{K_n}}]$. 
		% The matrix stacking the channels of all the UEs is $\bar{\G}\triangleq [\bar{\G}_1,\dots,\bar{\G}_N] \in \C^{M\times K}$, where $\bar{\G}_n\triangleq [\bar{\g}_{n_1},\dots,\bar{\g}_{n_{K_n}}]\in \C^{M\times K_n}$.
		
		% \begin{figure}[t!]
			% \centering
			% \includegraphics[width=0.5\textwidth]{FLiterationGLOBECOM2.eps}
			% \vspace{-8mm}
			% \caption{Detailed operation of one FL iteration of each group.}
			% \label{fig:time2}
			% \end{figure}
		
		% \begin{itemize}
			% \item \textbf{Downlink payload data transmission}:
			% \end{itemize}
		\textbf{Downlink payload data transmission}:
		The BS encodes the global training update intended for  UE $k$ of group $n$ into a symbol $s_{d,n_k}$, where $\EEE\{|s_{d,n_k}|^2\}=1$, and 
		% and use a ZF procoding scheme 
		% linear procoding scheme, i.e., MRT or ZF, 
		apply the ZF precoding vector 
		% $\uu_n = \sqrt{\frac{\eta_{n}}{M\hat{\sigma}_{n}^2}} \hat{\g}_{n}\,\,\text{for the MRT scheme}$ or 
		$\uu_{n_k}\! =\! \sqrt{\eta_{n_k}\hat{\sigma}_{n_k}^2(M\!-\!K_{total})} \hat{\G}(\hat{\G}^H\hat{\G})^{-1}\e_{n_k,K_{total}}
		% \,\,\text{for the ZF scheme},
		$ to precode the symbol,
		% \begin{align*}
			% &\z_n = \sqrt{\frac{\eta_{n}}{M\hat{\sigma}_n^2}} \hat{\g}_{n}\,\,\text{for the MRT scheme, or}
			% \\
			% &\z_n = \sqrt{\eta_{n}\hat{\sigma}_n^2(M-N)} \hat{\G}(\hat{\G}^H\hat{\G})^{-1}\e_{n,N}\,\,\text{for the ZF scheme},
			% \end{align*}
		where $\eta_{n_k}$ is a power control coefficient, 
		$\e_{n_k,K_{total}}$ is the $n_k$-th column of $\pmb{I}_{K_{total}}$, and $M \geq K_{total}$ is required.
		% Since the number of UEs is normally much larger than the number of groups, we allocate the same power for the UEs of each group to propose a low-complexity power control algorithm (which is presented in the next section). 
		The transmitted signal at the BS is thus given as $\x_{d}\!\!=\!\! \sqrt{\rho_{d}}\sum_{n'\in\NN}\sum_{\ell\in \K_{n'}}\uu_{n'_\ell} s_{d,n'}$, where $\rho_{d}$ is the maximum normalized transmit power at the BS. The transmitted power at the BS is required to meet the average normalized power constraint, i.e., $\EEE\{|x_{d}|^2\}\leq \rho_d$, which can be expressed through the following constraint:
		\begin{align}
			\label{powerdupperbound}
			\sum_{n\in\NN}\sum_{k\in\K_n}\eta_{n_k} \leq 1.
		\end{align}
		% The recevied signal at UE $k$ in group $n$ is
		% \begin{align}
			% \label{ydnk}
			% \nonumber
			% y_{d,n_k} 
			% &= \sum_{m\in\MM} g_{mn_k}x_{d,m} + w_{n_k}
			% \\
			% &= \sqrt{\rho_{d}}\sum_{m\in\MM}\sum_{n'\in\NN}\sum_{n'_\ell\in\K_{n'}} \sqrt{\eta_{mn'}}g_{mn_k}\hat{g}_{mn'_\ell}^*s_{d,n}+ w_{n_k}
			% \end{align}
		The achievable rate $R_{d,n_k}(\ETA)$ at UE $k$ of group $n$
		% and $R_{d,n_k}^{\text{ZF}}(\ETA)$ (bps) 
		% of UE $k$ of group $n$ under the MRT and ZF schemes are 
		is given 
		% respectively 
		as \cite[(10)]{sadeghi18TWC}
		% (shown at the top of the next page),
		% The data rate $R_{d,n}$ (bps) of sending the global training update from the APs to the UEs of group $n$ is thus expressed as
		% \begin{align}\label{R:d:multi}
			% R_{d,n}(\ETA) \! = \! \min_{k\in\K_n} R_{d,n_k}^{\text{multi}}(\ETA).
			% \end{align}
		% \begin{figure*}[t!]
			\vspace{-0mm}
			\begin{align}
				% \label{RdMRT}
				% &R_{d,n_k}(\ETA)\!=\!\frac{\tau_c-\tau_p}{\tau_c}B\log_2\!\Bigg(\!1+\!
				% \frac
				% {M \rho_d \sigma_{n_k}^2\eta_{n}}
				% {\rho_d \beta_{n_k} \sum_{n'\in\NN} \eta_{n'} 
					% +1}\Bigg)
				% R_{d,n_k}(\ETA) & = \frac{\tau_c-\tau_{d,p}}{\tau_c}B\log_2 \big(1+\text{SINR}_{d,n_k}(\ETA)\big)
				% \\
				\label{RdZF}
				% & R_{d,n_k}^{\text{ZF}}(\ETA)\!=\! \frac{\tau_c\!\!-\!\tau_p}{\tau_c}B\log_2\! \Bigg(\!\!1\!+\! \frac
				% {(M-N)\rho_d\sigma_{n_k}^2\eta_{n}}
				% {\rho_d (\beta_{n_k}\!\!-\!\sigma_{n_k}^2) \sum_{n'\in\NN} \eta_{n'}
					% \!+\!1}\!\!\Bigg),
				R_{d,n_k}(\ETA) &= \frac{\tau_c - \tau_{d,p}}{\tau_c}B\log_2 \big( 1 + \text{SINR}_{d,n_k}(\ETA)\big),
			\end{align}
			% \end{figure*}
		\vspace{-3mm}
		
		\noindent
		where $\ETA\triangleq\{\eta_{n_k}\}_{n\in\NN,k\in\K_n}$, $B$ is the bandwidth, 
		% $\text{SINR}_{d,n_k} (\ETA) = \frac{\Upsilon_{n_k}(\eta_n)}{ \Pi_{n_k}(\ETA)}$ 
		and $\text{SINR}_{d,n_k}(\ETA) =
		\frac{(M-K_{total})\rho_d \hat{\sigma}_{n_k}^2\eta_{n_k}}
		{\rho_d (\beta_{n_k} - \hat{\sigma}_{n_k}^2) \sum_{n'\in\NN}\sum_{n'_\ell\in\K_{n'}} \eta_{n'_\ell} +1}
		% \frac{\Upsilon_{n_k}(\eta_n)}{\Pi_{n_k}(\ETA)}
		$ is the effective downlink SINR\footnote{Although all the UEs of one group have the same encoded symbol, their achievable rates can be different (and hence, their transmissions do not finish simultaneously). This is feasible when using a code that sends a maximum number of parity bits corresponding to the UEs with the smallest SINR. Here, each UE will stop listening as soon as it successfully decodes its message. Thus, the UEs with higher SINRs can stop listening earlier than those with smaller SINRs.}.
		% where
		% $\Upsilon_{n_k}(\eta_n)=M \rho_d \hat{\sigma}_{n_k}^2\eta_{n}$, $\Pi_{n_k}(\ETA) = \rho_d \beta_{n_k} \sum_{n'\in\NN} \eta_{n'}+1$,
		% $\Upsilon_{n_k}(\eta_n)=(M-K_{total})\rho_d \hat{\sigma}_{n_k}^2\eta_{n_k}$, and $\Pi_{n_k}(\ETA) = \rho_d (\beta_{n_k} - \hat{\sigma}_{n_k}^2) \sum_{n'\in\NN}\sum_{\ell\in\K_{n'}} \eta_{n'} +1$.

		% \subsubsection{Downlink unicasting under unicast channel estimation}
		% The CPU encodes the global training update into different symbols $s_{d,n_k}$ for different UEs of each group $n$ and sends all the symbols to the APs. Here, the transmitted signal at AP $m$ with conjugate beamforming is $x_{d,m}\!\!=\!\! \sqrt{\rho_{d}}\sum_{n\in\NN}\sum_{k\in\K_n}\sqrt{\eta_{mn}}\hat{g}_{mn_k}^*s_{d,n_k}$. The $R_{d,n_k}$ (bps) the data rate of sending the global training update to UE $k$ of group $n$ is given in \eqref{Rduni} \cite[Theorem~1]{ngo17TWC}. 
		% \begin{figure*}[t!]
			% \begin{align}\label{Rduni}
				% R_{d,n_k}^{\text{uni}}(\ETA)\!=\!\frac{\tau_c\!-\!\tau_p^{\text{uni}}}{\tau_c}B\log_2\!\Bigg(1\!+\!\frac
				% {\rho_d\big(\sum_{m\in\MM}\eta_{mn}^{1/2}_{mn_k}^2\big)^2}
				% { \rho_d\sum_{n'\in\NN}\sum_{\ell\in\K_{n'}}\sum_{m\in\MM}\eta_{mn'}_{mn'_\ell}^2 \beta_{mn_k}
					% \!+\!1}\Bigg)
				% \end{align}
			% \hrulefill
			% \end{figure*}
		% % The transmission time from the APs to UE $k$ of group $n$ in the unicast scheme is given by
		% % \begin{align}\label{tduni}
			% % t_{d,n_k}^{\text{uni}}(\ETA) = \frac{S_{d,n}}{R_{d,n_k}^{\text{uni}}(\ETA)}.
			% % \end{align}
		
		% \begin{itemize}
			% \item \textbf{Downlink delay}:
			% \end{itemize}
		\textbf{Downlink delay}:
		Let $S_{d,n}$ (bits) be the data size of the global training update of group $n$. 
		The transmission time from the BS to UE $k$ of group $n$ is given by
		% under the MRT and ZF schemes are, respectively, given by
		\begin{align*}\label{}
			% t_{d,n_k}(\ETA) = \frac{S_{d,n}}{R_{d,n_k}(\ETA)}, \,\,
			t_{d,n_k}(\ETA) = \frac{S_{d,n}}{R_{d,n_k}(\ETA)}.
		\end{align*}
		
		% \begin{itemize}
			% \item \textbf{Energy consumption for downlink transmission at BS}:
			% \end{itemize}
		\textbf{Energy consumption for the downlink transmission}:
		Denote by $N_0$ is the noise power. The energy consumption  for transmitting the global update to the UE $k$ of group $n$ is the product of the transmit power $\rho_d N_0 \eta_{n_k}$ and the delay for the downlink transmission to this UE. Therefore, the total energy consumption at the BS for all groups is
		% the energy consumption 
		% under the MRT and ZF schemes are given respectively by
		\vspace{-0mm}
		\begin{align}
			\nonumber
			\!E_{d}(\ETA) 
			\!\!=\!\!\! \sum_{n\in\NN}\!\sum_{k\in\K_n}\!\! \rho_d N_0 \eta_{n_k} \!
			% \max_{n_k\in\K_n} 
			t_{d,n_k}(\ETA)
			% \\
			% &
			\!\!=\!\!\! \sum_{n\in\NN}\!\sum_{k\in\K_n}\!\! \rho_d N_0 \eta_{n_k}\!
			% \max_{n_k\in\K_n} 
			\frac{S_{d,n}}{R_{d,n_k}\!(\ETA)}.
			% \\
			% &\leq \rho_d N_0 \Big(\sum_{n\in\NN} \eta_n\Big) \max_{n\in\NN} \max_{n_k\in\K_n} t_{d,n_k}(\ETA)
			% \\
			% E_{d}^{\text{ZF}}(\ETA) & = \sum_{n\in\NN} \rho_d N_0 \eta_n \max_{n_k\in\K_n} \frac{S_{d,n}}{R_{d,n_k}^{\text{ZF}}(\ETA)}
		\end{align}
		
		\vspace{-0mm}
		\subsubsection{Step (S2)} After receiving the global update, each UE executes $L$ local computing rounds over its data set to compute its local update. 
		
		% \begin{itemize}
			% \item \textbf{Local computation}:
			% \end{itemize}
		\textbf{Local computation}:
		Let $c_{n_k}$ (cycles/sample) be the number of processing cycles for a UE $k$ to process one data sample \cite{tran19INFOCOM}. Denote by $D_{n}$ (samples) and $f_{n_k}$ (cycles/s) the size of the local data set and the processing frequency of the UE $k$ of group $n$, respectively. 
		% \textcolor{blue}{Let $L$ be the number of local computation iterations.} 
		The computation time at UE $k$ of group $n$ is then given by \cite{vu20TWC,tran19INFOCOM}
		\vspace{-0mm}
		\begin{align*}\label{}
			t_{C,n_k}(f_{n_k}) = \frac{LD_nc_{n_k}}{f_{n_k}}.
		\end{align*}
		% Given the limited computational resource at the UEs, we only focus on the delay of computing the local updates at the UEs. Since the computational resource of the CPU is much more abundant than that of the UEs, the latency of aggregating the global UL training updates at the CPU is negligible, and hence ignored.
		
		% \begin{itemize}
			% \item \textbf{Energy consumption for local computation at UEs}:
			% \end{itemize}
		\textbf{Energy consumption for local computation at the UEs}:
		The energy consumption at UE $k$ of group $n$ for computing its local training update is given as \cite{tran19INFOCOM,vu20TWC}
		\begin{align*}\label{}
			E_{C,n_k}(f_{n_k}) = L\frac{\alpha}{2}c_{n_k}D_nf_{n_k}^2,
		\end{align*}
		where $\frac{\alpha}{2}$ is the effective capacitance coefficient of the UEs' computing chipset.
		
		\vspace{-1mm}
		\subsubsection{Step (S3)}
		In this step, UEs' local updates are transmitted to the BS. 
		% This uplink transmission include two main phases (uplink channel estimation and uplink data transmission) per coherence block.
		
		% \begin{itemize}
			% \item \textbf{Uplink channel estimation}:
			% \end{itemize}
		\textbf{Uplink channel estimation}:
		In each coherence block, each UE sends its pilot of length $\tau_{u,p}$  to the BS. We assume that the pilots of all the UEs are pairwisely orthogonal, which requires the pilots of length
		$\tau_{u,p} \geq K_{total}$.
		% Let $\sqrt{\tau_{p}}\VARPHI_{k}\in\C^{\tau_{t}\times 1}$ be the pilot sequence transmitted from UE $k\in\NN$, where $\|\VARPHI_{k}\|^2\!\!=\!\!1,\forall k\in\K\triangleq \{1,\dots,K\}$.
		% Denote by $g_{mk} \!=\! (\beta_{mk})^{1/2}\tilde{g}_{mk}$ the channel from UE $k$ to AP $m$, where $\beta_{mk}$ and $\tilde{g}_{mk} \sim \CN(0,1)$ are the large-scale fading and small-scale fading channel coefficients, respectively. At AP $m$, $g_{mk}$ is estimated by using the received pilots and  the minimum mean-square error (MMSE) estimation. 
		The MMSE estimate $\bar{\g}_{n_k}$ of $\g_{n_k}$ is distributed according to $\CN(\pmb{0},\bar{\sigma}_{n_k}^2\pmb{I}_M)$, where
		$\bar{\sigma}_{n_k}^2 = \frac{\tau_{u,p} \rho_{p} \beta_{n_k}^2}{\tau_{u,p} \rho_{p}\beta_{n_k}+1}$ \cite{sadeghi18TWC}. 
		% The matrix stacking the channels of all the UEs is $\bar{\G}\triangleq [\bar{\G}_1,\dots,\bar{\G}_N]$, where $\bar{\G}_n\triangleq [\bar{\g}_{n_1},\dots,\bar{\g}_{n_{K_n}}]$.
		
		% \begin{remark}
			% Our scheme uses different pilots and different pilot assignment methods for the downlink transmission of Step (S1) and the uplink transmission of this step, which is more flexible 
			% \end{remark}
		
		% \begin{itemize}
			% \item \textbf{Uplink payload data transmission}:
			% \end{itemize}
		\textbf{Uplink payload data transmission}:
		After computing the local update, UE $k$ of group $n$ encodes this update into symbols denoted by $s_{u,n_k}$, where $\EEE\{|s_{u,n_k}|^2\}=1$, and sends baseband signal $x_{u,{n_k}}\!=\!\sqrt{\rho_{u}\zeta_{n_k}}s_{u,{n_k}}$ to the BS, where $\rho_{u}$ is the maximum normalized transmit power at each UE and $\zeta_{n_k}$ is a power control coefficient.
		This signal is subjected to the average transmit power constraint, i.e., $\EEE\left\{|x_{u,n_k}|^2\right\}\leq \rho_u$, which is can be expressed in a per-UE constraint as
		\begin{align}\label{poweruupperbound}
			\zeta_{n_k}\leq 1,\forall n\in\NN,n_k\in \K_n.
		\end{align}
		
		After receiving data from all UEs, the BS uses the estimate channels and ZF scheme to detect the UEs' message symbols. The ZF precoder requires $M \geq K_{total}$. The achievable rate
		% $R_{u,{n_k}}(\ZETA)$ and
		% $R_{u,{n_k}}^{\text{ZF}}(\ZETA)$ 
		(bps) of UE $k$ in group $n$ is given by \cite[(3.28)]{ngo16}
		% under the MRC and ZF schemes are given respectively as \cite[(3.41), (3.28)]{ngo16}
		% \eqref{RuMRT} and \eqref{RuZF}, 
		% (shown at the top of the next page). 
		\begin{align}
			% \label{RuMRT}
			% R_{u,n_k}(\ZETA)
			% &= \frac{\tau_c-\tau_{u,p}}{\tau_c}B \log_2 \big( 1 + \text{SINR}_{u,n_k}(\ZETA)\big)
			% \\
			\label{RuZF}
			R_{u,n_k}(\ZETA)
			&= \frac{\tau_c-\tau_{u,p}}{\tau_c}B \log_2 \big( 1 + 
			\text{SINR}_{u,n_k}(\ZETA)
			\big),
		\end{align}
		\vspace{-2mm}
		
		\noindent
		where
		% $\text{SINR}_{u,n_k} (\ZETA) = \frac{\Psi_{n_k}(\zeta_n)}{\Xi_{n_k}(\ZETA)}$ and
		$\text{SINR}_{u,n_k} (\ZETA) \triangleq
		\frac{(M-K_{total}) \rho_u \bar{\sigma}_{n_k}^2 \zeta_{n_k}} {\rho_u \sum_{n' \in \NN} \sum_{n'_\ell \in \K_{n'}}  (\beta_{n'_\ell} - \bar{\sigma}_{n'_\ell}^2) \zeta_{n'_\ell} + 1}
		% \frac{\Psi_{n_k}(\zeta_n)}{\Xi_{n_k}(\ZETA)}
		$ is the effective uplink SINR.
		% where
		% $\Psi_{n_k}(\zeta_{n_k}) = M \rho_u \bar{\sigma}_{n_k}^2 \zeta_{n_k}$, $\Xi_{n_k}(\ZETA) = \rho_u \sum_{n' \in \NN} \sum_{n'_\ell \in \K_{n'}} \bar{\sigma}_{n'_\ell}^2 \zeta_{n'_\ell} + 1$,
		% $\Psi_{n_k}(\zeta_{n_k}) = (M-K_{total}) \rho_u \bar{\sigma}_{n_k}^2 \zeta_{n_k}$, and $\Xi_{n_k}(\ZETA) = \rho_u \sum_{n' \in \NN} \sum_{n'_\ell \in \K_{n'}}  (\beta_{n'_\ell} - \bar{\sigma}_{n'_\ell}^2) \zeta_{n'_\ell} + 1$.   
		
		% \begin{itemize}
			% \item \textbf{Uplink delay}:
			% \end{itemize}
		\textbf{Uplink delay}:
		Denote by $S_{u,n}$ (bits) the data size of the local training update of group $n$.
		The transmission time from UE $k$ of group $n$ to the BS is given by
		\begin{align*}\label{}
			% t_{u,n_k}(\ZETA) = \frac{S_{u,n}}{R_{u,n_k}(\ZETA)}, \,\,
			t_{u,n_k}(\ZETA) = \frac{S_{u,n}}{R_{u,n_k}(\ZETA)}.
		\end{align*}
		
		% \begin{figure*}[t!]
			% \begin{align}\label{SINRuMRT}
				% \nonumber
				% & R_{u,n_k}^{\text{multi}}(\ZETA)
				% =\frac{\tau_c-\tau_p^{\text{multi}}}{\tau_c}B \times
				% \\
				% & \quad \log_2\Bigg(1+ \frac
				% {\rho_u\zeta_{n}\left(\sum_{m\in\MM}\sigma_{mn_k}^2\right)^2}
				% { \sum_{n_\ell\in\K_{n} \setminus n_k} \zeta_{n_\ell} \big( \sum_{m\in\MM} \sigma_{mn_k} \sigma_{mn_\ell} \big)^2
					% +
					% \rho_u\sum_{n'\in\NN}  \sum_{n'_\ell\in\K_{n'}} \zeta_{n'_\ell} \sum_{m\in\MM} \sigma_{mn_k}^2 \beta_{mn'_\ell}
					% +\sum_{m\in\MM}\sigma_{mn_k}^2}\!
				% \Bigg)
				% \\
				% \label{RuZF}
				% &R_{u,n_k}^{\text{ZF}}(\ZETA)=\frac{\tau_c-\tau_p}{\tau_c}B\log_2\Bigg(1+
				% \frac{ (M-N)\rho_u \sigma_{n_k}^2 \zeta_{n_k}}
				% { (M-N)\rho_u \sum_{n_\ell\in\K_n \setminus n_k} \sigma_{n_\ell}^2 \zeta_{n_\ell} + 
					% \rho_u \sum_{n'\in\NN} \sum_{\ell\in\K_{n'}} (\beta_{mn'_\ell} - \sigma_{mn_k}^2)\zeta_{n'_\ell}
					% +1}
				% \Bigg)
				% \end{align}
			% \hrulefill
			% \end{figure*}
		
		% \begin{itemize}
			% \item \textbf{Energy consumption for uplink transmission at UEs}:
			% \end{itemize}
		\textbf{Energy consumption for the uplink transmission}:
		The energy consumption for the uplink transmission at a UE is the product of the uplink power and the transmission time. In particular, the energy consumption at UE $k$ of group $n$ 
		% under the MRC and ZF schemes are given respectively as
		is given as \cite{tran19INFOCOM,vu20TWC}
		\vspace{-1mm}
		\begin{align*}
			% E_{u,n_k}(\ZETA) &= \rho_u N_0\zeta_{n_k} t_{u,n_k}(\ZETA) = \frac{\rho_u N_0\zeta_{n_k}S_{u,n}} {R_{u,n_k}^{\text{MRC}}(\ZETA)}.
			% \\
			E_{u,n_k}(\ZETA) &= \rho_u N_0\zeta_{n_k} t_{u,n_k}(\ZETA) = \frac{\rho_u N_0\zeta_{n_k}S_{u,n}} {R_{u,n_k}(\ZETA)}.
		\end{align*}
		
		\vspace{-1mm}
		\begin{remark}
			We obtain the achievable downlink and uplink rates in \eqref{RdZF} and \eqref{RuZF}, respectively, under the case that all users participate in the transmission. However, as shown from the two proposed schemes in Fig.~\ref{fig:time1}, at a particular time, some UEs may have finished their transmission, and thus, do not participate in the downlink or uplink transmission with other UEs at the same time. This will not cause any issue with our design because the rates \eqref{RdZF} and \eqref{RuZF} are still always achievable under this case.
		\end{remark}
		
		\subsubsection{Step (S4)}
		After receiving all the local updates, the BS computes its global update. Since the computational capability of the central server is much more powerful than those of the UEs, the delay of computing the global update is negligible.

		\vspace{-1mm}
		\section{Problem Formulation and Solution}
		\label{sec:PF}
		\vspace{-1mm}
		In practice, different groups are likely to start their FL processes at different times and have different number of FL iterations depending on their learning targets. Therefore, minimizing the energy consumption of the whole FL processes of all groups at the same time is tremendously difficult due to complicated synchronization among all groups. Instead, we aim at minimizing the total energy consumption in one FL iteration for all groups, which also leads to the total energy consumption reduction of the whole FL processes of all groups.
		
		\vspace{-2mm}
		\subsection{Asynchronous Scheme}
		\vspace{-1mm}
		% As shown in Fig.~\ref{fig:time}, at the current iteration, every group needs to wait until all groups finishing their iterations to start a new iteration. Therefore, the time of one FL iteration of every group is the longest delay caused by the most straggler UE in the network. 
		
		The problem of minimizing the total energy consumption of one FL iteration for all groups is formulated as follows.
		\begin{subequations}\label{Pmain}
			\begin{align}
				\label{CFPmulti}
				\!\!\!\!\!\underset{\ETA,\f,\ZETA}{\min} 
				& E_{total} \triangleq\! E_{d}(\ETA)\! +\!\!\! \sum_{n\in\NN}\!\sum_{n_k\in\K_n} (E_{C,n_k}(f_{n_k}) \!+\! E_{u,n_k}(\ZETA))
				\\
				\nonumber
				\!\!\!\!\!\mathrm{s.t.}\,\,
				&
				\eqref{powerdupperbound}, \eqref{poweruupperbound}
				\\
				\label{powerlowerbound}
				& 0\leq \eta_{n}, 0\leq \zeta_{n_k}, \forall n,n_k
				\\
				\label{fbound}
				& 0 \leq f_{n_k} \leq f_{\max}, \forall n,n_k
				\\
				\label{QoSbound}
				& t_{d,n_k}(\ETA) + t_{C,n_k}(f_{n_k}) + t_{u,n_k}(\ZETA) 
				\leq t_{\text{QoS}}, \forall n,n_k
				\\
				\label{syncbound}
				& \max_{n\in\NN}\max_{n_k\in\K_n}\!\! t_{d,n_k}
				\!\leq\! \min_{n\in\NN}\min_{n_k\in\K_n} \big(t_{d,n_k} \!+\! t_{C,n_k}\big),
			\end{align}
		\end{subequations}
		\noindent
		where $\f\triangleq\{f_{n_k}\}_{n\in\NN,n_k\in\K_n} $. Here, \eqref{QoSbound} guarantees the execution time of one FL iteration below a threshold $t_{\text{QoS}}$ for maintaining the quality of service, and 
		\eqref{syncbound} is introduced to ensure that all the UEs send their local update during the uplink mode of the BS. The right-hand side of \eqref{syncbound} models the first UE that finishes its downlink transmission and local computation, while the left-hand side presents the slowest UE finishes its downlink transmission as seen in Fig.~1(a).
		% This constraint models the scenario that the first UE which finishes its downlink transmission and local computation starts its uplink transmission after . 
		% Finding the globally optimal solution of \eqref{P:multi} is challenging. This paper instead aims to propose a solution approach that is suitable for practical implementation.
		
		To solve \eqref{Pmain}, we rewrite it in the following more tractable epigraph form
		\begin{subequations}\label{Pmainepi}
			\begin{align}
				\label{CF:shortP:epi}
				\nonumber
				\!\!\!\!\!\!\underset{\x}{\min} \,\,
				& \widetilde{E}_{total} \triangleq \sum_{n\in\NN}\sum_{n_k\in\K_n} \rho_d N_0 S_{d,n} \omega_{n_k} 
				\\
				& + \sum_{n\in\NN}\!\sum_{n_k\in\K_n} \!\!\!\Big( L\frac{\alpha}{2}c_{n_k}D_nf_{n_k}^2 \!\!+\!\! \rho_u N_0\theta_{n_k}S_{u,n} \Big)
				\\
				\mathrm{s.t.}\,\,
				\nonumber
				&
				\eqref{powerdupperbound}, \eqref{poweruupperbound}, \eqref{powerlowerbound}, \eqref{fbound}
				\\
				\label{Rdlowerbound}
				& r_{d,n_k}\leq R_{d,n_k} (\ETA), \forall n,n_k
				\\
				\label{Rulowerbound}
				& r_{u,n_k}\leq R_{u,n_k} (\ZETA), \forall n,n_k
				\\
				\label{rlowerbound}
				&  0 \leq r_{d,n_k}, 0 \leq r_{u,n_k}, \forall  n,n_k
				\\
				\label{tdgroupbound}
				& 
				% \frac{\eta_{n_k}}{r_{d,n_k}} \leq \omega_{n_k}, \forall n, n_k
				\eta_{n_k}\leq r_{d,n_k} \omega_{n_k}, \forall n, n_k
				\\
				\label{tubound}
				& 
				% \frac{\zeta_{n_k}}{r_{u,n_k}}  \leq \theta_{n_k}, \forall n,n_k 
				\zeta_{n_k}\leq r_{u,n_k}  \theta_{n_k}, \forall n,n_k
				\\
				\label{tbound}
				& \frac{S_{d,n}}{r_{d,n_k}} + \frac{LD_nc_{n_k}}{f_{n_k}} + \frac{S_{u,n}}{r_{u,n_k}} \leq t_{\text{QoS}}, \forall n, n_k
				% \\
				% \label{syncbound2a}
				% & b \leq q
				\\
				\label{syncbound2a}
				& 
				% \frac{S_{d,n}}{r_{d,n_k}} \leq q, \forall n,n_k
				S_{d,n}\leq r_{d,n_k} q, \forall n,n_k
				\\
				\label{syncbound2b}
				& q \leq q_{1,n_k} + q_{2,n_k}, \forall n,n_k
				\\
				\label{syncbound2c}
				& 0 \leq q_{1,n_k}, 0 \leq q_{2,n_k}, \forall n,n_k
				\\
				\label{syncbound2d}
				& 
				% q_{1,n_k} \leq \frac{S_{d,n}}{r_{d,n_k}}, \forall n,n_k
				q_{1,n_k} r_{d,n_k} \leq S_{d,n} , \forall n,n_k
				\\
				\label{syncbound2e}
				& q_{2,n_k} f_{n_k} \leq LD_nc_{n_k} , \forall n,n_k,
			\end{align}
		\end{subequations}
		where $\x \triangleq \{\ETA,\f,\ZETA,\rr_d,\rr_u,\OOmega,\THeta, q, \q_1, \q_2\}$, $\rr_d,\rr_u,\OOmega,\THeta,q,\q_1,\\ \q_2$ are additional variables,  $\rr_d=\{r_{d,n_k}\}$, $\rr_u=\{r_{u,n_k}\}$, $\OOmega=\{\omega_{n_k}\}$, $\THeta=\{\theta_{n_k}\}$,
		$\q_1=\{q_{1,n_k}\}, \q_2=\{\q_{2,n_k}\}, \forall n\in\NN,n_k\in\K_n$. Here,
		% \eqref{powerdbound} and \eqref{abound} represent \eqref{powerdupperbound}, while
		\eqref{syncbound2a}--\eqref{syncbound2e} come from \eqref{syncbound}.
		If we let
		$\vv \triangleq \{v_{n_k}\}$ and $\uu\triangleq \{u_{n_k}\}, \forall n\in\NN,n_k\in\K_n,$ with $v_{n_k}\triangleq \eta_{n_k}^{1/2}, \,\, u_{n_k} \triangleq \zeta_{n_k}^{1/2}, \forall n,n_k,$
		% \begin{align} \label{variabletransform}
			% v_{n_k}\triangleq \eta_{n_k}^{1/2}, \,\, u_{n_k} \triangleq \zeta_{n_k}^{1/2}, \forall n,n_k,
			% \end{align}
		then problem \eqref{Pmainepi} will be equivalent to
		\begin{subequations}\label{Pmainepiequi}
			\begin{align}
				\label{CF:shortP:epi}
				\!\!\!\!\!\!\underset{\widetilde{\x}}{\min} \,\,
				& \widetilde{E}_{total} 
				\\
				\mathrm{s.t.}\,\,
				\nonumber
				&
				\eqref{fbound}, \eqref{rlowerbound}, \eqref{tbound}-\eqref{syncbound2c}
				\\
				\label{Rdlowerbound2}
				& r_{d,n_k}\leq R_{d,n_k} (\vv), \forall n,n_k
				\\
				\label{Rulowerbound2}
				& r_{u,n_k}\leq R_{u,n_k} (\uu), \forall n,n_k
				\\
				\label{tdgroupbound2}
				& v_{n_k}^2 - r_{d,n_k}\omega_{n_k} \leq 0, \forall n, n_k
				\\
				\label{tubound2}
				& u_{n_k}^2 - r_{u,n_k} \theta_{n_k} \leq 0, \forall n,n_k 
				\\
				\label{powerdupperbound2}
				& \sum_{n\in\NN} \sum_{k\in\K_n} v_{n_k}^2 \leq 1
				\\
				\label{poweruupperbound2}
				& u_{n_k}^2\leq 1, \forall n,n_k
				\\
				\label{powerlowerbound2}
				& 0\leq v_{n_k}, 0\leq u_{n_k},\forall n,n_k,
				\\
				\label{syncbound2e2}
				& q_{1,n_k} r_{d,n_k} - S_{d,n}\leq 0, \forall n,n_k
				\\
				\label{syncbound2f2}
				& q_{2,n_k} f_{n_k} - LD_{n}c_{n_k}\leq 0, \forall n,n_k,
			\end{align}
		\end{subequations}
		where $\widetilde{\x}\triangleq\{\x,\vv,\uu\}\setminus\{\ETA,\ZETA\}$. Here, \eqref{tdgroupbound2} and \eqref{tubound2} follow from \eqref{tdgroupbound} and \eqref{tubound}, while \eqref{powerdupperbound2}--\eqref{poweruupperbound2} follow from \eqref{powerdupperbound}, \eqref{poweruupperbound}, and \eqref{powerlowerbound}. Problem \eqref{Pmainepiequi} is still difficult to solve due to nonconvex constraints \eqref{Rdlowerbound2}, \eqref{Rulowerbound2}, \eqref{tdgroupbound2}, \eqref{tubound2}, \eqref{syncbound2e2}, and \eqref{syncbound2f2}.
		
		To deal with these constraints, we first observe that the rates $R_{d,n_k}(\vv)$ and $R_{u,n_k}(\uu)$ of nonconvex constraints \eqref{Rdlowerbound2} and \eqref{Rulowerbound2} have the following concave lower bounds \cite[(20)]{nguyen17TCOM}:
		\begin{align}\label{Rdconcave} 
			\nonumber
			&\widetilde{R}_{d,n_k}(\vv) \triangleq
			\frac{\tau_c-\tau_{cp}}{\tau_c\log 2}B\Big[\log \Big(1 + \frac{(\Upsilon_{n_k}^{(i)})^2}  {\Pi_{n_k}^{(i)}} \Big)
			-\frac{(\Upsilon_{n_k}^{(i)})^2}{\Pi_{n_k}^{(i)}}
			\\
			&+2\frac{\Upsilon_{n_k}^{(i)}\Upsilon_{n_k}}{\Pi_{n_k}^{(i)}}
			-\frac{(\Upsilon_{n_k}^{(i)})^2(\Upsilon_{n_k}^2+\Pi_{n_k})}{\Pi_{n_k}^{(i)}((\Upsilon_{n_k}^{(i)})^2+\Pi_{n_k}^{(i)})}\Big]
			\leq R_{d,n_k}(\vv),
			\\
			\nonumber
			&\widetilde{R}_{u,n_k}(\uu) \triangleq
			\frac{\tau_c-\tau_{dp}}{\tau_c\log 2} B \Big[ \log\Big(1+\frac{(\Psi_{n_k}^{(i)})^2}{\Xi_{n_k}^{(i)}}\Big)
			-\frac{(\Psi_{n_k}^{(i)})^2}{\Xi_{n_k}^{(i)}}
			\\
			&+ 2\frac{\Psi_{n_k}^{(i)}\Psi_{n_k}}{\Xi_{n_k}^{(i)}}
			- \frac{(\Psi_{n_k}^{(i)})^2(\Psi_{n_k}^2+\Xi_{n_k})}{\Xi_{n_k}^{(i)}((\Psi_{n_k}^{(i)})^2+\Xi_{n_k}^{(i)})} \Big]
			\leq R_{u,n_k}(\uu),
		\end{align}
		\noindent
		where $\Pi_{n_k}(\vv) = \rho_d (\beta_{n_k} - \hat{\sigma}_{n_k}^2) \sum_{n'\in\NN}\sum_{n'_\ell\in\K_{n'}} v_{n'_\ell}^2 +1$,
		$\Upsilon_{n_k} (v_{n_k}) = \sqrt{(M-K_{total})\rho_d} \hat{\sigma}_{n_k}v_{n_k}$,
		% \begin{align}\label{Ruconcave}
			% \nonumber
			% &\widetilde{R}_{u,n_k}(\ETA) \triangleq
			% \frac{\tau_c-\tau_{dp}}{\tau_c}B\Bigg[\log_2 \Big(1 + \frac{\Psi_{n_k}^{(i)}}{\Xi_{n_k}^{(i)}} \Big) + 2\frac{\Psi_{n_k}^{(i)}}{(\Psi_{n_k}^{(i)} + \Xi_{n_k}^{(i)})}
			% \\
			% &\!\!- \frac{(\Psi_{n_k}^{(i)})^2}{(\Psi_{n_k}^{(i)} + \Xi_{n_k}^{(i)})\Psi_{n_k}} \!-\!\frac{\Psi_{n_k}^{(i)}\Xi_{n_k}}{(\Psi_{n_k}^{(i)} + \Xi_{n_k}^{(i)})\Xi_{n_k}^{(i)}}\Bigg]
			% \!\leq\! R_{d,n_k}(\ETA),
			% \end{align}
		$\Xi_{n_k}(\uu) = \rho_u \sum_{n' \in \NN} \sum_{n'_\ell \in \K_{n'}}  (\beta_{n'_\ell} - \bar{\sigma}_{n'_\ell}^2) u_{n'_\ell}^2 + 1$, and 
		$\Psi_{n_k} = \sqrt{(M-K_{total}) \rho_u} \bar{\sigma}_{n_k} u_{n_k}$.
		Next, the functions in the left-hand sides of constraints \eqref{tdgroupbound2}, \eqref{tubound2}, \eqref{syncbound2e2}, and \eqref{syncbound2f2} have the following convex upper bounds \cite{vu20TWC}:
		\vspace{-0mm}
		\begin{align}
			\nonumber
			&\!\!\!\!v_{n_k}^2\! -\! r_{d,n_k}\omega_{n_k} \!\leq\! h_{1,n_k}(v_{n_k},r_{d,n_k},\omega_{n_k})
			\triangleq 0.25
			% \frac{1}{4}
			\big[4v_{n_k}^2 \!+\!(r_{d,n_k} \!-
			\\
			&\!\!\!\! \omega_{n_k})^2 - 2(r_{d,n_k}^{(i)}\!+\!\omega_{n_k}^{(i)})(r_{d,n_k}\! +\! \omega_{n_k})\! + \!(r_{d,n_k}^{(i)}\!+\!\omega_{n_k}^{(i)})^2\big],
			\\
			\nonumber
			&\!\!\!\!u_{n_k}^2 \!-\! r_{u,n_k}\theta_{n_k} \!\leq\! h_{2,n_k}(u_{n_k},r_{u,n_k},\theta_{n_k})\triangleq 0.25
			% \frac{1}{4}
			\big[4u_{n_k}^2 \!+\! (r_{u,n_k}\!- 
			\\
			&\!\!\!\! \theta_{n_k})^2 - 2(r_{u,n_k}^{(i)}\!+\!\theta_{n_k}^{(i)})(r_{u,n_k} \!+\! \theta_{n_k}) \!+\! (r_{u,n_k}^{(i)}\!+\!\theta_{n_k}^{(i)})^2\big]
			\\
			\nonumber
			&\!\!\!\! q_{1,n_k} r_{d,n_k} \!-\! S_{d,n} \!\leq\!
			h_{3,n_k}(q_{1,n_k},r_{d,n_k})\triangleq 0.25
			% \frac{1}{4}
			\big[  (q_{1,n_k}\!+ r_{d,n_k})^2
			\\
			&\!\!\!\! \!\! -\! 2(q_{1,n_k}^{(i)}\!\!-\!r_{d,n_k}^{(i)}) (q_{1,n_k} \!\!-\! r_{d,n_k}) \!+\! (q_{1,n_k}^{(i)}\!\!-\!r_{d,n_k}^{(i)})^2 \!-\! 4S_{d,n}\big]
			\\
			\nonumber
			&\!\!\!\! q_{2,n_k} f_{n_k} \!-\! LD_{n}c_{n_k} \!\leq\!
			h_{4,n_k}(q_{2,n_k},f_{n_k})\triangleq  0.25
			% \frac{1}{4}
			\big[  (q_{2,n_k}\!+ f_{n_k})^2
			\\
			&\!\!\!\! \!\! -\! 2(q_{2,n_k}^{(i)}\!\!-\!f_{n_k}^{(i)}) (q_{2,n_k} \!\!-\! f_{n_k}) \!+\! (q_{2,n_k}^{(i)}\!\!-\!\!f_{n_k}^{(i)})^2 \!-\! 4LD_{n}c_{n_k}\big].
		\end{align}
		% The functions in the right-hand sides of the nonconvex constraints \eqref{syncbound2e} and \eqref{syncbound2f} have the following concave lower bounds \cite[(78)]{nguyen21TCOM}:
		% \begin{align}\label{qupperbound}
			% &h_{3,n_k}(r_{d,n_k})\triangleq
			% S_{d,n}\Bigg(\frac{2}{r_{d,n_k}^{(i)}} - \frac{r_{d,n_k}}{(r_{d,n_k}^{(i)})^2}\Bigg) \leq \frac{S_{d,n}}{r_{d,n_k}},
			% \\
			% &h_{4,n_k}(f_{n_k})\triangleq LD_nc_{n_k} \Bigg(\frac{2}{f_{n_k}^{(i)}} - \frac{f_{n_k}}{(f_{n_k}^{(i)})^2}\Bigg) \leq  \frac{LD_nc_{n_k}}{f_{n_k}}.
			% \end{align}
		
		As such, constraints \eqref{Rdlowerbound2}, \eqref{Rulowerbound2}, \eqref{tdgroupbound2}, \eqref{tubound2}, \eqref{syncbound2e2}, and \eqref{syncbound2f2} can now be approximated respectively by the following convex constraints 
		\vspace{-2mm}
		\begin{align}
			\label{Rdlowerboundapprox}
			&r_{d,n_k}  \leq \widetilde{R}_{d,n_k}(\vv), \forall n,n_k
			\\
			\label{Rulowerboundapprox}
			&r_{u,n_k}  \leq \widetilde{R}_{u,n_k}(\uu), \forall n,n_k
			\\
			\label{tubounddapprox}
			&h_{1,n_k}(v_{n_k},r_{d,n_k},\omega_{n_k})  \leq 0, \forall n,n_k
			\\
			\label{tuboundd2approx}
			&h_{2,n_k}(u_{n_k},r_{u,n_k},\theta_{n_k})  \leq 0, \forall n,n_k
			\\
			\label{syncbound2eapprox}
			&h_3(q_{1,n_k},r_{d,n_k}) \leq 0, \forall n,n_k
			\\
			\label{syncbound2fapprox}
			&h_4(q_{2,n_k},f_{n_k}) \leq 0, \forall n,n_k.
		\end{align}
		\vspace{-4mm}
		
		\noindent
		At iteration $(i+1)$, for a given point $\widetilde{\x}^{(i)}$, problem \eqref{Pmainepi} can finally be approximated by the following convex problem:
		%\begin{subequations}
			\begin{align}\label{Pmainepiapprox}
				\underset{\widetilde{\x}\in\widetilde{\FF}}{\min} \,\,
				& \widetilde{E}_{total},
			\end{align}
			%\end{subequations}
		where $\widetilde{\FF}\triangleq\!\{
		\eqref{fbound}, 
		\eqref{rlowerbound},\eqref{tbound}-\eqref{syncbound2c},
		\eqref{powerdupperbound2}-\eqref{powerlowerbound2},
		\eqref{Rdlowerboundapprox}-\eqref{syncbound2fapprox}
		\}$ is a convex feasible set.
		
		In Algorithm~\ref{alg}, we outline the main steps to solve problem \eqref{Pmain}.
		Let $\FF\triangleq\{
		\eqref{powerdupperbound}, \eqref{poweruupperbound},
		\eqref{powerlowerbound} - 
		\eqref{syncbound}\}$ be the feasible set of \eqref{Pmain}.
		Starting from a random point $\widetilde{\x}\in\FF$, we solve \eqref{Pmainepiapprox} to obtain its optimal solution $\widetilde{\x}^*$, and use $\widetilde{\x}^*$ as an initial point in the next iteration. The algorithm terminates when an accuracy level of $\varepsilon$ is reached. In the case when $\widetilde{\FF}$ satisfies Slater's constraint qualification condition, Alg.~\ref{alg} will converges to a Karush-Kuhn-Tucker solution of \eqref{Pmainepi} (hence \eqref{Pmain}) \cite[Theorem 1]{Marks78OR}. In contrast, Alg.~\ref{alg} will converges to a Fritz John solution of \eqref{Pmainepi} (hence \eqref{Pmain}). 
		
		\begin{algorithm}[!t]
			\caption{Solving problem \eqref{Pmain}}
			\begin{algorithmic}[1]\label{alg}
				\STATE \textbf{Initialize}: Set $i\!=\!0$ and choose a random point $\widetilde{\x}^{(0)}\!\in\!\FF$.
				\REPEAT
				\STATE Update $i=i+1$
				\STATE Solving \eqref{Pmainepiapprox} to obtain its optimal solution $\widetilde{\x}^*$
				\STATE Update $\widetilde{\x}^{(i)}=\widetilde{\x}^*$
				\UNTIL{convergence}
			\end{algorithmic}
			\vspace{+0mm}
			\textbf{Output}: $(\ETA^*,\ZETA^*,\f^*)$
		\end{algorithm}

		\vspace{-3mm}
		\subsection{Synchronous Scheme}
		\vspace{-2mm}
		The optimization problem of this scheme is formulated as
		\begin{subequations}\label{Pmainsyn}
			\begin{align}
				\label{CFPmulti}
				\!\!\!\!\!\underset{\ETA,\f,\ZETA}{\min} \,\,
				& E_{d}(\ETA) + \!\!\sum_{n\in\K_n}\!\sum_{k\in\K_n} \!\!(E_{C,n_k}(f_{n_k})\! +\! E_{u,n_k}(\ZETA))
				\\
				\nonumber
				\mathrm{s.t.}\,\,
				&
				\eqref{powerdupperbound}, \eqref{poweruupperbound}, \eqref{powerlowerbound}, \eqref{fbound}
				\\
				\nonumber
				\label{QoSboundsyn}
				&\max_{n\in\NN}\max_{n_k\in\K_n}t_{d,n_k}(\ETA) + \max_{n\in\NN}\max_{n_k\in\K_n}t_{C,n_k}(\f)
				\\
				&\qquad+ \max_{n\in\NN}\max_{n_k\in\K_n}t_{u,n_k}(\ZETA) 
				\leq t_{\text{QoS}}.
			\end{align}
		\end{subequations}
		Here, constraint \eqref{QoSboundsyn} captures the nature of ``step-by-step'', i.e., every UE needs to wait for the UEs of all groups to finish one step before starting the next step as seen in Fig.~\ref{fig:time1}(b). Compared to \eqref{QoSboundsyn}, \eqref{QoSbound} provides more flexibility for allocating times of Steps (S1)--(S3) for each UE since the UEs do not need to wait for other UEs to start a new step. 
		
		Using the similar procedure to solve problem \eqref{Pmain} above, we approximate \eqref{Pmainsyn} by the following convex problem
		\begin{subequations}\label{Pmainsyncepiapprox}
			\begin{align}
				\underset{\widehat{\x}}{\min} \,\,
				& \widetilde{E}_{total} 
				\\
				\nonumber
				\mathrm{s.t.}\,\,
				& \eqref{fbound}, 
				\eqref{rlowerbound},
				\eqref{powerdupperbound2}-\eqref{powerlowerbound2},
				\eqref{Rdlowerboundapprox}-\eqref{tuboundd2approx}
				\\
				& t_d + t_C + t_u \leq t_{\text{QoS}}
				\\
				& 
				\frac{S_{d,n}}{r_{d,n_k}} \leq t_d, \forall n,n_k
				% S_{d,n} \leq r_{d,n_k}  t_d , \forall n,n_k
				\\
				& 
				\frac{LD_nc_{n_k}}{f_{n_k}} \leq t_C, \forall n,n_k
				% LD_nc_{n_k}  \leq f_{n_k} t_C , \forall n,n_k
				\\
				& \frac{S_{u,n}}{r_{u,n_k}}  \leq  t_u , \forall n,n_k,
			\end{align}
		\end{subequations}
		where $\rr_d,\rr_u,\OOmega,\THeta,t_d,t_C,t_u$ are additional variables and $\widehat{\x} \triangleq \{\vv,\f,\uu,\rr_d,\rr_u,\OOmega,\THeta,t_d,t_C,t_u\}$.
		Then, problem \eqref{Pmainsyn} can be solved using Algorithm~\ref{alg} for iteratively solving \eqref{Pmainsyncepiapprox}.
		\vspace{-2mm}
		\subsection{Complexity Analysis}
		\vspace{-1mm}
		% For comparison purposes, we assume that $K_n = K, \forall n$. 
		Problem \eqref{Pmainepiapprox} can be transformed to an equivalent problem that involves $V_1\triangleq (9K_{total}+1)$ real-valued scalar variables, $L_1\triangleq (8K_{total}+4)$ linear constraints, $Q_1\triangleq 11K_{total}$ quadratic constraints. Therefore, problem \eqref{Pmainepiapprox} requires a complexitiy of $\OO(\sqrt{L_1+Q_1}(V_1+L_1+Q_1)V_1^2)$ \cite{tam16TWC}. The transformed version of problem \eqref{Pmainsyncepiapprox} involves a smaller number of variables and constraints than the version of problem \eqref{Pmainepiapprox}, i.e, $V_2\triangleq (6K_{total}+3)$ real-valued scalar variables, $L_2\triangleq (6K_{total}+1)$ linear constraints, $Q_2\triangleq 5K_{total}$ quadratic constraints. Therefore, problem \eqref{Pmainsyncepiapprox} has the complexity of $\OO(\sqrt{L_2+Q_2}(V_2+L_2+Q_2)V_2^2)$ which is lower than that of problem \eqref{Pmainepiapprox}. 
		As such, it is expected that the synchronous scheme requires a lower complexity than the asynchronous scheme. However, the synchronous scheme requires more signaling overhead to achieve synchronization than the asynchronous scheme.

		% \begin{remark}
			% Let (A,B) be the case of using a precoding scheme A at Step (S1) and a decoding scheme B at Step (S3). 
			% The optimization problems for the other cases, i.e.,  (ZF, MRC), (MRT, ZF) and (ZF, ZF), are formulated similarly to that for (MRT, MRC) above because they share the same mathematical structure. Therefore, our proposed Alg.~\ref{alg} can be applied straighforwardly to solve the formulated problems of those cases. 
			% \end{remark}
		\vspace{-1mm}
		\section{Numerical Examples}
		\vspace{-1mm}
		\label{sec:sim}
		\subsection{Network Setup and Parameter Setting}
		\vspace{-1mm}
		Consider a mMIMO network in a square of $D\times D$ km$^2$ where the BS is at the center and the UEs are randomly located. 
		% The distances between adjacent APs are at least $50$ m. 
		We set $\tau_c\!=\!200$ samples.
		The large-scale fading coefficients, i.e., $\beta_{mn_k}$, are modeled in the same manner as \cite[Eqs. (37), (38)]{emil20TWC}.
		% To estimate channels, a random pilot assignment is used as in \cite{vu20TWC}.
		For ease of presentation, we assume that all groups have the same number of UEs, i.e., $K_n =K, \forall n$. The total number of UEs is thus $NK$.
		We choose 
		% $\tau_{cp} \!=\! N$, 
		$\tau_{d,p} =\tau_{u,p} \!=\! NK$, $S_d\!=\!S_u\!=\!20$ MB, noise power $\sigma_0^2\!=\!-92$ dBm, $L=50$, $f_{\max}=4 \times 10^9$ cycles/s, $D_n = 5\times 10^6$ samples, $c_{n_k} = 20$ cycles/samples \cite{tran19INFOCOM}, for all $n,n_k$, $\alpha=5\times 10^{-30}$, $t_{\text{QoS}} = 5$ s. 
		Let $\tilde{\rho}_d\!=\!6$ W, $\tilde{\rho}_u\!=\!0.2$ W and $\tilde{\rho}_p\!=\!0.2$ W be the maximum transmit power of the APs, UEs and uplink pilot sequences, respectively. The maximum transmit powers $\rho_d$, $\rho_u$ and $\rho_p$ are normalized by the noise power. 
		
		%\begin{figure}[]
			%  \centering
			%  \vspace{-0mm}
			%  {\includegraphics[width=0.31\textwidth]{ConvC2.pdf}\label{subfig:1b}}
			%  \vspace{-3mm}
			%  \caption{The convergence of Algorithm~\ref{alg:main} ($M\!=\!10, N\!=\!6,N_{\text{QoL}}\!=\!3,K\!=\!3$).}
			%  \label{Fig:1}
			%  \vspace{-0mm}
			%\end{figure}
		\vspace{-1mm}
		\subsection{Results and Discussions}
		\vspace{-1mm}
		Note that there are no other existing works studying wireless networks for supporting multiple FL groups. Therefore, to evaluate the effectiveness of our proposed asynchronous scheme (\textbf{OPT\_Async}) and synchronous scheme (\textbf{OPT\_Sync}), we consider the following heuristic schemes:
		\begin{itemize}
			% \item \textbf{BL1\_CF}: The downlink power allocated to all groups are the same, i.e., $\eta_{mn}\sum_{n_k\in \K_n}\sigma_{mn_k}^2 = 1/N, \forall n$. The transmitted power of each UE is $\eta_{n_k}=1, \forall n,k$. The processing frequency of UEs are optimized. 
			\item \textbf{Heuristic\_Async} (Heuristic solution for asynchronous scheme): The downlink power to the UEs of all groups are the same, i.e., $\eta_{n_k}\!=\!\frac{1}{NK}$ and the transmitted power of each UE is $\eta_{n_k}=1, \forall n, n_k$. The processing frequencies are $f_{n_k} = \frac{LD_nc_{n_k}}{t_{\text{QoS}-t_{d,n_k} - t_{u,n_k}}}, \forall n,n_k$. 
			\item \textbf{Heuristic\_Sync} (Heuristic solution for synchronous scheme):  
			Similar to Heuristic\_Async except for the processing frequencies which are set as $f_{n_k} = \frac{LD_nc_{n_k}}{t_{\text{QoS}-\max_{n\in\NN}\max_{n_k\in\K_n}t_{d,n_k} - \max_{n\in\NN}\max_{n_k\in\K_n}t_{u,n_k}}}, \forall n,n_k$. 
		\end{itemize} 
		
		\begin{figure}[t!]
			\centering
			\vspace{-0mm}
			{\includegraphics[width=0.38\textwidth]{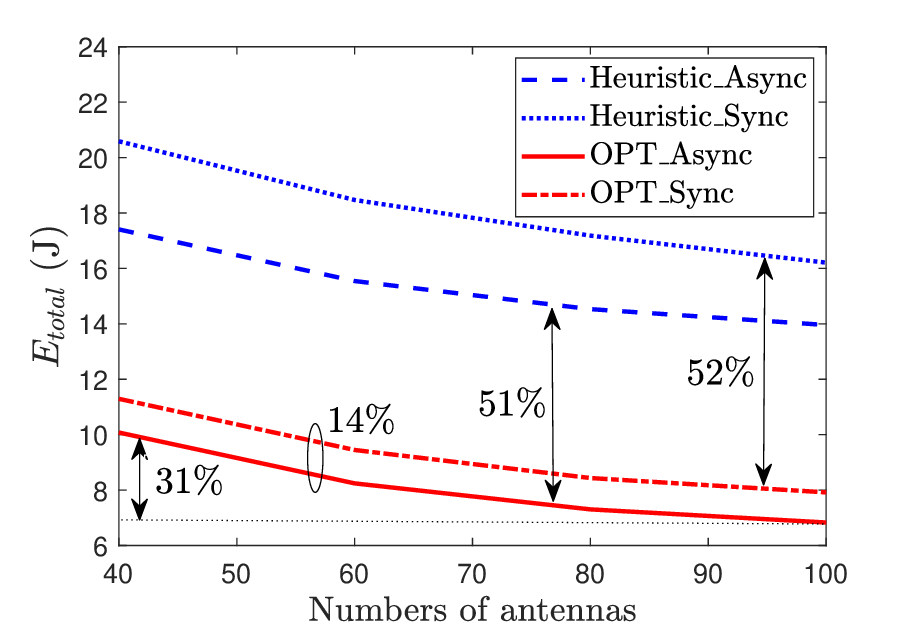}\label{fig:a}}
			\caption{Comparison among the proposed approach and baselines ($K=10$ (users per group), $N\!=\!3$ groups, $D=0.25$ km).}
			\label{Fig:sim1}
			\vspace{-5mm}
		\end{figure}
		
		\begin{figure}[t!]
			\centering
			{\includegraphics[width=0.39\textwidth]{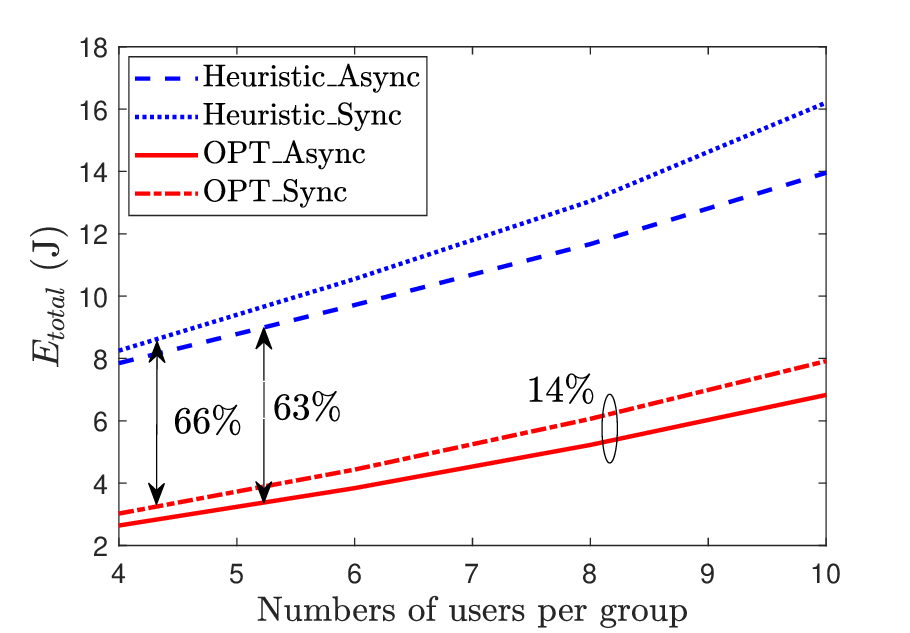}\label{fig:b}}
			\vspace{-0mm}
			\caption{Comparison among the proposed approach and baselines ($M = 100$ (antennas), $N\!=\!3$ groups, $D=0.25$ km).}
			\label{Fig:sim2}
			\vspace{-0mm}
		\end{figure}
		
		Figs.~\ref{Fig:sim1} and~\ref{Fig:sim2} compare the total energy consumption of one FL iteration among the considered schemes. As seen, our proposed schemes give the best performance. Specifically, compared to heuristic schemes, the energy reduction are up to $52\%$  with $M=100$, $K=10$, and up to $66\%$ with $M=100$, $K=4$. The figures not only demonstrate the significant advantage of a joint allocation of power and processing frequency, but also show the benefit of using massive MIMO to support FL. Thanks to massive MIMO technology, the data rate of each UE increases when the number of antennas increases, leading to lower delays and then a decrease of $31\%$ in the total energy consumption of one FL iteration as shown in Fig.~\ref{Fig:sim1}. 
		% It can be seen from Fig ... that when the UEs increases, the time of one iteration increase due to stronger mutual interference. 
		
		Figs.~\ref{Fig:sim1} and~\ref{Fig:sim2} also shows that the asynchronous scheme slightly outperforms the synchronous scheme. In particular, the energy reduction in one FL iteration is up to only $14\%$ with $M=100$, $K=10$. This is reasonable because the UEs in the asynchronous scheme do not need to wait for other UEs. As such, they have more time resource, and hence, can save more energy by using lower processing frequencies than those in the synchronous scheme. However, minimizing energy consumption results in maximizing the lowest data rate. Therefore, data rates obtained by the asynchronous scheme are relatively similar to those by the synchronous scheme, which leads to a similar performance of both schemes.  
		\vspace{-1mm}
		\section{Conclusion}
		\vspace{-1mm}
		\label{sec:con}
		This work has proposed two novel schemes with mMIMO as energy-efficient solutions for future wireless networks to support multiple FL groups. Using successive convex approximation techniques, we have also successfully proposed an algorithm to allocate power and processing frequency in order to minimize the energy consumption in each  FL iteration. Numerical results showed that our proposed schemes significantly reduces the energy consumption of each FL iteration compared to heuristic schemes. They also confirmed that in terms of energy savings, the asynchronous scheme is a better choice to support multiple FL groups than the synchronous scheme, though at the cost of higher complexity. 
		\vspace{-1mm}

		\ifCLASSOPTIONcaptionsoff
		\newpage
		\fi

		\vspace{-0mm}
		\section*{Acknowledgment}
		\vspace{-0mm}
		The work of T.~T.~Vu and H.~Q.~Ngo was supported by the U.~K. Research and Innovation Future Leaders Fellowships under Grant MR/S017666/1. The work of Erik~G.~Larsson was supported in part by ELLIIT and the Knut and Alice Wallenberg Foundation. The work of Minh~N.~Dao was partially supported by Federation University Australia under Grant RGS21-8.
		
		\vspace{-2mm}
		\begin{spacing}{1}
			\bibliographystyle{IEEEtran}
			\bibliography{IEEEabrv,newidea2021}
			% \bibliography{newidea2021}
		\end{spacing}
		
	\end{document}